\documentclass[runningheads]{llncs}
\usepackage[T1]{fontenc}

\usepackage{graphicx}
\usepackage{amsmath,amssymb}
\usepackage[nolist]{acronym}
\usepackage{cite}
\usepackage{hyperref}
\usepackage{placeins}
\usepackage{subfigure}
\usepackage{color}

\begin{document}
\title{Interpreting Latent Spaces of Generative Models for Medical Images using Unsupervised Methods}
\titlerunning{Interpreting Latent Spaces of Generative Models}
%
\author{ Julian Schön\inst{1,2} \and Raghavendra Selvan\inst{1,3} \and
Jens Petersen\inst{1,2}}
\authorrunning{J. Schön et al.}
%
\institute{Department of Computer Science, University of Copenhagen, Denmark \\\email{\{julian.e.s, raghav, phup\}@di.ku.dk} \and Department of Oncology, Rigshospitalet, Denmark\and
Department of Neuroscience, University of Copenhagen, Denmark}
\maketitle              
\begin{abstract}
Generative models such as Generative Adversarial Networks (GANs) and Variational Autoencoders (VAEs) play an increasingly important role in medical image analysis. The latent spaces of these models often show semantically meaningful directions corresponding to human-interpretable image transformations. However, until now, their exploration for medical images has been limited due to the requirement of supervised data. Several methods for unsupervised discovery of interpretable directions in GAN latent spaces have shown interesting results on natural images. This work explores the potential of applying these techniques on medical images by training a GAN and a VAE on thoracic CT scans and using an unsupervised method to discover interpretable directions in the resulting latent space. We find several directions corresponding to non-trivial image transformations, such as rotation or breast size. Furthermore, the directions show that the generative models capture 3D structure despite being presented only with 2D data. The results show that unsupervised methods to discover interpretable directions in GANs generalize to VAEs and can be applied to medical images. This opens a wide array of future work using these methods in medical image analysis. The code and animations of the discovered directions are available online at \url{https://github.com/julschoen/Latent-Space-Exploration-CT}.
\keywords{Generative models\and unsupervised learning\and interpretability\and CT.}
\end{abstract}
\section{Introduction}
The combination of deep learning and medical images has emerged as a promising tool for diagnostics and treatment. One of the main limitations is the often small dataset sizes available for deep learning. Generative models can be used to mitigate this by synthesizing and augmenting medical images \cite{Kazeminia20}.\\
\acp{GAN} \cite{Goodfellow14} have emerged as the prominent generative model for image synthesis. Consequently, research focusing on the interpretability of \acp{GAN} has unfolded. At their inception, Radford et al. \cite{Radford16} showed meaningful vector arithmetic in the latent space of \acp{DCGAN}. For several years, the methods used for discovering interpretable directions in latent spaces have been supervised \cite{Goetschalk19, Plumerault20, Jahanian20} or based on simple vector arithmetic \cite{Radford16}. Especially in medical image analysis, supervision is expensive as it typically involves radiologists or other experts' time. Recently, several unsupervised methods for discovering interpretable directions in \ac{GAN} latent spaces were proposed \cite{Voynov20, Harkonen20, Shen21}. Due to being unsupervised, they seem more promising for the medical domain. However, it is still unclear if they work with the often more homogeneous images and the smaller dataset sizes encountered in this field.\\
Next to \acp{GAN}, the interpretability of \acp{VAE} \cite{Kingma14VAE} has also been studied extensively. However, the investigation has mainly focused on obtaining disentangled latent space representations \cite{Higgins17, Kim18}. While this shows promising results, it might not be possible without introducing inductive biases \cite{Locatello19}. Applying the approaches for the unsupervised discovery of interpretable directions in latent spaces developed for \acp{GAN} to \acp{VAE} might yield an alternative route for the investigation of interpretability in \acp{VAE}. Thus, if the same methods that have shown promising results on \acp{GAN} are effective on \acp{VAE}, then \acp{VAE} can be trained without restrictions on the latent space, therefore not incorporating inductive biases while still having the benefit of interpretability and explicit data approximation.\\
{\bf Contributions}: We employ a technique for the unsupervised discovery of interpretable directions in the latent spaces of \acp{DCGAN} and \acp{VAE} trained on \ac{CT} scans. We show that these methods used to interpret the latent spaces of \acp{GAN} generalize to \acp{VAE}. Further, our results provide insights into the applicability of these methods for medical image analysis. We evaluate the directions obtained and show that non-trivial and semantically meaningful directions are encoded in the latent space of the generative models under consideration. These directions include both transformations specific to our dataset choice and ones that likely generalize to other data. In particular, this allows for future work considering semantic editing of medical images in latent spaces of generative models.

\section{Background}
\subsection{Generative Latent Models}
As the backbone of this work we use generative latent models. We employ two of the most popular model types in \acp{GAN} \cite{Goodfellow14} for implicit and \acp{VAE} \cite{Kingma14VAE} for explicit approximation of the data distribution \cite{goodfellow2016nips}. 

Given the discriminator $D$, the generator $G$, the latent distribution $p_z$, the data distribution $p_{data}$, and binary cross-entropy as the loss the \ac{GAN} optimization is given by:\\
\begin{equation}
    \underset{G}{min}\:\underset{D}{max}\: V(D,G) = \mathbb E_{x\sim p_{data}} [\log D(x)]+ \mathbb E_{z\sim p_z}[\log (1-D(G(z)))].
\end{equation}

We optimize the \ac{VAE} using the \ac{ELBO} with additional scaling factor $\beta$ \cite{Higgins17} given by:
\begin{equation}
    \mathcal{L}_{VAE} = -\mathbb E_{q_\theta} [\log\: p_{\phi} (x|z)] + \beta D_{KL}[q_\theta (z|x) || p(z)]
\end{equation}
where the first term is referred to as the reconstruction loss, with $p_\phi$ giving the likelihood parameterised by $\phi$, and the second term as the regularization loss given by the \ac{KLD}, with $q_\theta$ giving the approximate posterior parameterised by $\theta$ and $p(z)$ is the prior given by $p(z)\sim \mathcal{N}(0,I)$.\\

\subsection{Discovery of Interpretable Directions in Latent Spaces}
Several unsupervised methods to find interpretable directions in \ac{GAN} latent spaces have been proposed \cite{Voynov20, Harkonen20, Shen21}. In Härkönen et al; Shen et al. \cite{Harkonen20, Shen21} the directions are orthogonal. This constraint is relaxed in Voynov and Babenko \cite{Voynov20}. As interpretable directions do not have to be orthogonal, we employ the method suggested by Voynov and Babenko \cite{Voynov20}. The proposed method can be applied to any pretrained latent generative model $G$. The objective is to learn distinct directions in the latent space of $G$ by learning a matrix $A$ containing directions and a reconstructor $R$ to distinguish between them. Since $A$ and $R$ are learned jointly, the directions of $A$ are likely to be interpretable, semantically meaningful, and affect all images equally. Otherwise, distinguishing between the directions would be hard, and consequently, the accuracy of $R$ would suffer.\\
\begin{figure}[h!]
    \centering
    \includegraphics[width=\textwidth]{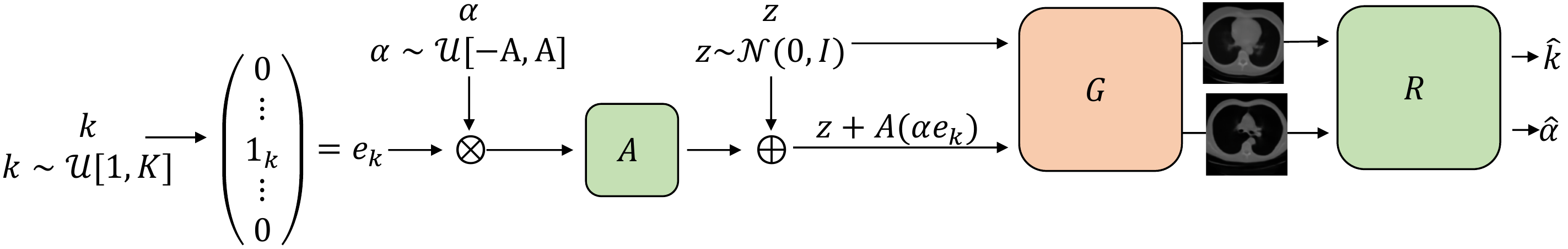}
    \caption{Schematic overview of the learning protocol suggested by Voynov and Babenko. The upper path corresponds to the original latent code $z\sim\mathcal N(0,I)$ and the lower path corresponds to the shifted code $z+A(\alpha e_k)$ (Adapted from \cite{Voynov20}).}
    \label{fig:dir_arch}
\end{figure}

Formally, the method learns a matrix $A\in\mathbb{R}^{d\times K}$, where $d$ is the dimensionality of the latent space of $G$, and $K$ is the number of directions that will be discovered. Thus, the columns of $A$ correspond to discovered directions and are optimized during the training process to be easily distinguishable. Further, let $z\sim\mathcal N(0,I)$ be a latent code, $e_k$ an axis-aligned unit vector with $k\in[1,...,K]$ and $\alpha$ a scalar. Then, we can define the image pair $(G(z), G(z+A(\alpha e_k)))$ where $G(z)$ is the original image generated by latent code $z$ and $G(z+A(\alpha e_k))$ is a shifted image from the original latent code $z$ shifted along the $k$th discovered direction by amount $\alpha$. Thus, $\alpha$ is a 'knob' controlling the magnitude of the shift. Given such an image pair, the method optimizes the reconstructor $R$ presented with that pair to predict the shift direction $k$ and amount $\alpha$.  Figure \ref{fig:dir_arch} illustrates the architecture. The optimization objective is given by:
\begin{equation}
    \label{eq:dir_loss}
    \underset{A,R}{min} \: \mathbb E_{z,k,\alpha} [L_{cl}(k,\hat{k}) + \gamma L_s(\alpha, \hat{\alpha})]
\end{equation}
where $k$ and $\alpha$ are the direction and amount respectively, and $\hat{k}$ and $\hat{\alpha}$ are the predictions. The classification term $L_{cl}$ is given by cross-entropy. Further, we can use the classification term to get the \ac{RCA}, i.e., the accuracy of predicting the direction. Finally, the shift term $L_s$ is given by the mean absolute error, and the regularization factor $\gamma$.

\section{Material \& Methods}
\subsection{Data}
We use \ac{LIDC-IDRI} \cite{Armato11} provided by \ac{TCIA}. It consists of clinical thoracic \ac{CT} scans of $1010$ patients collected from diagnostic and lung cancer screenings and is assembled by seven academic centers and eight medical imaging companies. We consider each axial slice as an individual image. Thus, our dataset consists of $246,016$ \ac{CT} slices. We resize the images to $128\times 128$ pixels to limit computational demands and limited the data to a range of $[-1000,2000]Hu$ to reduce the amount of outlier values and normalized using min-max scaling.

\subsection{Models \& Training}
Since this study focuses on the potential of unsupervised exploration of latent spaces for medical images, we use simple generative models. We use a \ac{DCGAN} based on Radford et al. \cite{Radford16}, improving training stability by introducing one-sided label smoothing \cite{Salimans16}, replacing the fixed targets $1$ of the real labels with smoothed values randomly chosen from the interval $[0.9,1]$. Additionally, we add $0$-mean and $0.1$ standard deviation Gaussian noise to the discriminator input \cite{Arjovsky17}, incrementally reducing the standard deviation and finally removing it at the midpoint of training. The encoder and decoder of the \ac{VAE} are based on ResNet \cite{Kaiming16}, and we use $\beta=0.01$ to improve reconstruction quality. For both generative models, we use a latent space size of $d=32$ as it showed the best trade-off between image quality and compactness of the latent space. We refer to the provided GitHub repository for implementation details. We train the \ac{GAN} and the \ac{VAE} for $50$ epochs selecting the best weights out of the last $5$ by considering the models \ac{FID} \cite{Heusel17} on test data. We use binary cross-entropy as loss for the \ac{GAN} and log mean squared error \cite{Yu20} as reconstruction loss for the \ac{VAE}. We use Adam \cite{Kingma14Adam} with a learning rate of $0.0002$ and $0.0001$ to optimize the \ac{GAN} and \ac{VAE}, respectively. The best model weights yield a \ac{FID} of $33.4$ for the \ac{GAN} and $93.9$ for the \ac{VAE} on the test data.\\
To find interpretable latent directions, we use two different reconstructor architectures, based on LeNet \cite{LeCun98} and ResNet18. We experiment with $A$ having unit length or orthonormal columns as suggested by Voynov and Babenko \cite{Voynov20}. We set the number of directions $K$ equal to the size of the latent space, i.e., $K=32$, and experiment with increasing it to $K=100$. We observe significantly faster convergence when using the ResNet reconstructor. Thus, when using $K=32$, we train the model for $25,000$ iterations using LeNet and $3,000$ iterations using the ResNet reconstructor. When $K=100$, we train the \ac{VAE} for $75,000$ and $4,000$ iterations with the LeNet and ResNet reconstructors respectively. For the \ac{GAN} we observe slower convergence. Thus, we train for $250,000$ and $10,000$ iterations with the LeNet and ResNet reconstructors, respectively. Since we cannot have $K>d$ for orthonormal directions, we only use $A$ with columns of unit length for $K=100$. We evaluate direction models using the \ac{RCA} and the shift loss $L_s$ from Equation \ref{eq:dir_loss}. Further, we follow the ablation provided by Voynov and Babenko \cite{Voynov20} and use a regularization factor $\gamma = 0.25$. To evaluate the directions, preliminary labeling was done by the first author with eight animations, each showing different latent vectors per direction. Next, each direction and preliminary label was considered on eight static images. The evaluator does not have formal training in medical image interpretation, and it is possible that more experienced evaluators could have discovered more interesting directions.

\section{Experiments \& Results}
\begin{figure}[h!]
    \centering
    \includegraphics[width=\textwidth]{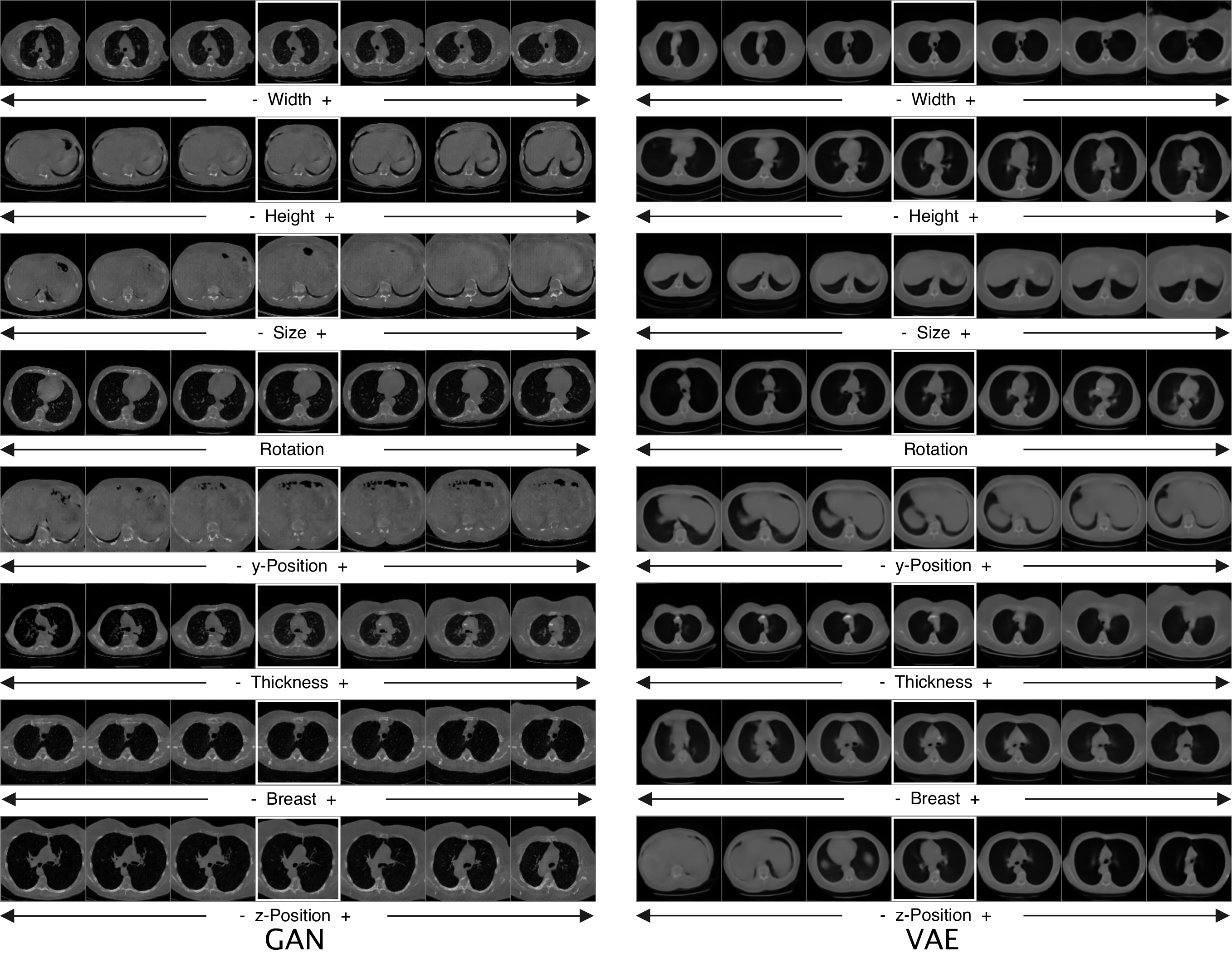}
    \caption{Interpretable directions using $A^{32\times 100}$ with unit length columns, LeNet as reconstructor, and the \ac{GAN} and \ac{VAE} as generative models. The central images correspond to the original latent vector. The left/right images correspond to shifts.}
    \label{fig:main}
\end{figure}

We perform several experiments to investigate the unsupervised exploration of latent spaces of deep generative models. First, we train using orthonormal di- rections and directions of unit length. We also experiment with increasing the number of directions. Finally, we perform all experiments both with a \ac{DCGAN} and a \ac{VAE} as generative models. All results are obtained without supervision, except the labeling of the selected directions. The \ac{RCA} and Ls of the differ- ent experiments are presented in Table \ref{tab:RCALs}. 
\begin{table}[h]
  \caption{Reconstructor Classification Accuracy (RCA) and $L_s$ for all model configurations for ResNet and LeNet as reconstructor.}
  \label{tab:RCALs}
  \centering
  \begin{tabular}{c||c|c|c|c|c|c}
     & \multicolumn{2}{c|}{\textbf{Orthogonal}} & \multicolumn{2}{c|}{\textbf{Unit Length}} & \multicolumn{2}{c}{\textbf{100 Directions}}\\
    & RCA & $L_s$ & RCA & $L_s$ & RCA & $L_s$\\
    \hline
    \bfseries GAN ResNet & $0.9236$ & $0.2538$ & $0.9383$ & $0.1949$ & $0.9522$ & $0.1560$ \\
    \bfseries GAN LeNet & $0.8559$ & $0.3317$ & $0.9062$ & $0.2439$ & $0.9305$ & $0.1406$ \\
    \bfseries VAE ResNet & $0.9939$ & $0.1040$ & $0.9947$ & $0.1086$ & $0.9861$ & $0.1117$ \\
    \bfseries VAE LeNet & $0.9800$ & $0.1421$ & $0.9895$ & $0.1090$ & $0.9791$ & $0.0962$
  \end{tabular}
  
\end{table}
We observe that the \ac{VAE} always outperforms the \ac{GAN} with respect to both \ac{RCA} and $L_s$. Further, using directions of unit length achieves higher \ac{RCA} than orthonormal directions and lower $L_s$ in all but one case. We also observe higher \ac{RCA} when using ResNet over LeNet as a reconstructor. In contrast, LeNet achieves a lower $L_s$ when $K$ is set to $100$.\\
Voynov and Babenko \cite{Voynov20} mention that a larger $K$ does not harm interpretability but alleviates entanglement and may lead to more duplicate directions. We observe the same behavior with $K=100$ as opposed to $K=32$.\\
Our results show eight consistent directions: width, height, size, rotation, $y$-position, thickness, breast size, and $z$-Position. All model configurations find all eight directions with varying degrees of entanglement. In this work, we omit directions entangled to such a degree that there is no clear interpretation dominating the image transformation. Thus, all configurations find at least a subset of the directions above in a sufficiently disentangled manner. We present animations of all discovered directions in the provided GitHub repository.
Figure \ref{fig:main} shows all eight directions for the \ac{VAE} and \ac{GAN}. The directions presented are obtained using LeNet as reconstructor and $K=100$. Directions obtained using different model configurations are presented in Appendix \ref{app:results_all}.
Our results show that enforcing orthonormal directions increases entanglement. Finally, we observe that when using a LeNet reconstructor, more of the obtained directions are easily interpretable compared to using a ResNet reconstructor.
\section{Discussion}
In this work, we explored the latent spaces of deep generative models to discover semantically meaningful directions. We next elaborate on some of the findings of our experiments.\\
{\bf Influence of $\boldsymbol{K}$}: We observe less entanglement when increasing $K$. Thus, we hypothesize that lower $K$ likely makes the reconstructor classification task easier, as there are less classes, lessening the need for disentanglement. If so, when increasing $K$, the increasing classification difficulty forces the model to disentangle the directions more.\\
{\bf Orthonormal Directions}: While constraining the directions to be orthonormal still leads to the same subset of interpretable directions being discovered, their quality suffers. This aligns with the observations of Voynov and Babenko \cite{Voynov20}. However, their results show that some datasets benefit from orthonormal directions, leading to more interesting directions. We do not observe this on our data, and the lack of disentanglement is also clear from the lower \ac{RCA} of the methods using orthonormal directions. Thus, it seems likely that directions offering semantic meaning are not necessarily orthonormal, strengthening our reasoning for choosing this method over  Härkönen et al; Shen et al. \cite{Harkonen20, Shen21}.\\
{\bf Choice of Reconstructor}: When $K=32$ both reconstructors show similar qualitative results, more entangled directions, $L_s$ is larger, and ResNet quantitatively outperforms LeNet. For $K=100$, LeNet produces better qualitative results than ResNet. This is also evident in the quantitative results with LeNet and $K=100$ achieving the lowest $L_s$. While ResNet has a higher \ac{RCA}, \ac{RCA} gives a measure of duplicate directions and only partially describes interpretability. Since LeNet performed best when using $K=100$ and the increased number of directions benefited disentanglement, we prefer LeNet as reconstructor.\\
{\bf Consistent Discovery of Interpretable Directions}: The same subset of human interpretable directions appears for all models with varying degrees of entanglement. Recent work has shown non-linear directions to be less entangled \cite{Tzelepis21} which could be studied further. The directions are validated by showing that the same set is discovered in the latent space of both the generative models. The resulting directions we discover show non-trivial image transformations. In particular, the directions changing the $z$-Position of the latent vector demonstrates that the models learn the 3D structure of the data despite being trained on 2D images. While the focus of discovering directions in latent spaces has mainly been on \acp{GAN} in recent years, we see that the same methods apply to \acp{VAE}. Since \acp{VAE} allow for explicit data approximation, they have a practical benefit over \acp{GAN} when considering the usefulness of these methods.\\
{\bf Impact \& Applications}: Improving interpretability of \acp{GAN} and \acp{VAE} is important and addressed in this work by finding and visualizing meaningful latent space directions and providing novel insights into the learned representations. The method is shown to generalize to \acp{VAE}, indicating that the latent spaces of \acp{VAE} and \acp{GAN} can be interpreted in similar ways. However, shorter convergence times on the \ac{VAE} when learning the directions indicate that VAEs latent spaces could be inherently easier to interpret. Unsupervised exploration further benefits the medical image domain due to the lack of well-supervised datasets, and more importantly, it could lead to surprising results outside of what we are explicitly supervising methods to find.\\
Our work can further be used for context-aware image augmentation and editing. Image augmentation using synthetic data improves downstream machine learning tasks on medical images \cite{Chlap21} and can alleviate both the small dataset sizes and imbalance inherent to medical imaging \cite{Kazeminia20, Yi19}. Our results could be used to explore more diverse augmentations, e.g., adjusting for sex and weight imbalances. Additionally, our work might offer an alternate unsupervised approach to disease-aware image editing \cite{Saboo21}.\\
We see further applications needing more investigation, such as exploring the potential in consistency regularization and multi-modal datasets. For example, finding directions corresponding to adding or removing contrast in scans. Further, the approach we use has been shown to be effective in unsupervised saliency detection and segmentation on natural images \cite{Voynov21, Melas21, Voynov20}.\\
{\bf Limitations}: The main limitations we observe in our work are based on the methodology for unsupervised exploration. First, while the RCA and shift loss give some insights into convergence, the implications of overfitting need to be investigated. In particular, deciding how many training iterations to use is difficult as model performance can not be assessed on independent data. Further, the lack of evaluation metrics makes the choice of reconstructor difficult. We tried to mitigate this by using \ac{RCA} and $L_s$ for quantitative and human interpretation for qualitative analysis. Nevertheless, further investigation is needed to find good evaluation metrics. Second, the large amount of resulting directions makes evaluation difficult and time-consuming. This is particularly challenging in medical image analysis as evaluation may involve trained evaluators such as radiologists. Further automation or introducing a hierarchy of interpretability could be a focus of future work. Next to the methodological limitations, we see further potential for expanding our work to 3D generative models and more datasets in the future.

\section{Conclusion}
In this work, we have demonstrated for the first time that techniques for unsupervised discovery of interpretable directions in the latent space of generative models yield good results on medical images. While the interpretability of latent spaces is arguably an abstract concept depending on those interpreting, our results show that generative models learn non-trivial, semantically meaningful directions when trained on \ac{CT} images of the thorax. We encounter directions with the same semantic meaning regardless of the generator or direction discovery model, indicating a general structure of the latent spaces. Further, our results show that the generative models' latent spaces capture the 3D structure of the \ac{CT} scans despite only being trained on 2D slices. The work opens up the possibility of exploring these techniques for unsupervised medical image segmentation, interpolation, augmentation, and more.

\subsubsection{Acknowledgements} The authors acknowledge the National Cancer Institute and the Foundation for the National Institutes of Health, and their critical role in the creation of the free publicly available LIDC/IDRI Database used in this study. The authors would like to thank Anna Kirchner and Arnau Morancho Tardà for help in preparation of the manuscript. Jens Petersen is partly funded by research grants from the Danish Cancer Society (grant no. R231-A13976) and Varian Medical Systems.

%
%
\bibliographystyle{splncs04}
\bibliography{mybibliography}
\clearpage
\appendix
\section{Discovered Directions for all Model Configurations.}
\label{app:results_all}
\begin{figure}[h]
  \centering
    \subfigure{
      \label{fig:VaeLeOrtho}
      \includegraphics[width=0.39\linewidth]{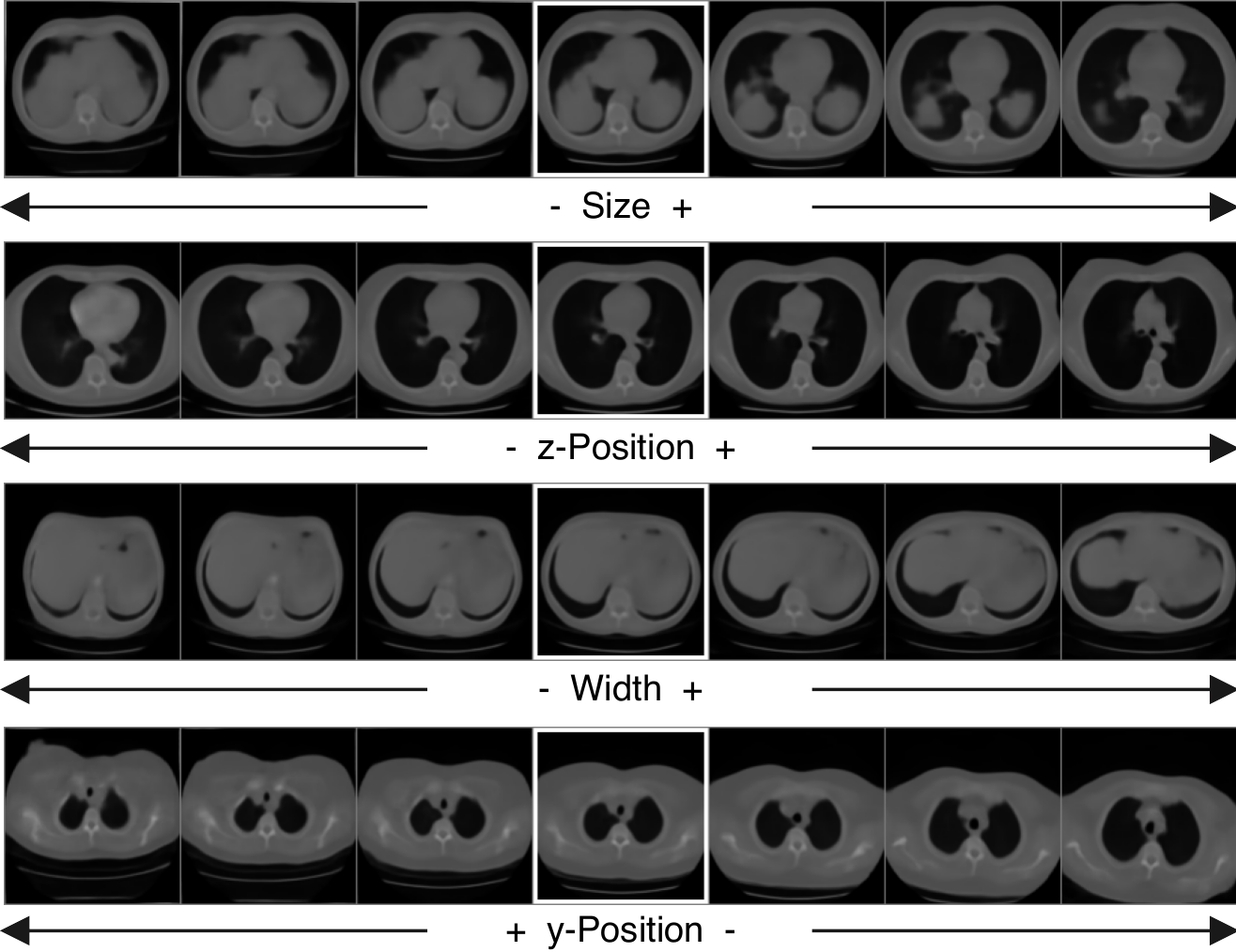}
    }\qquad 
    \subfigure{
      \label{fig:GanLeOrtho}
      \includegraphics[width=0.39\linewidth]{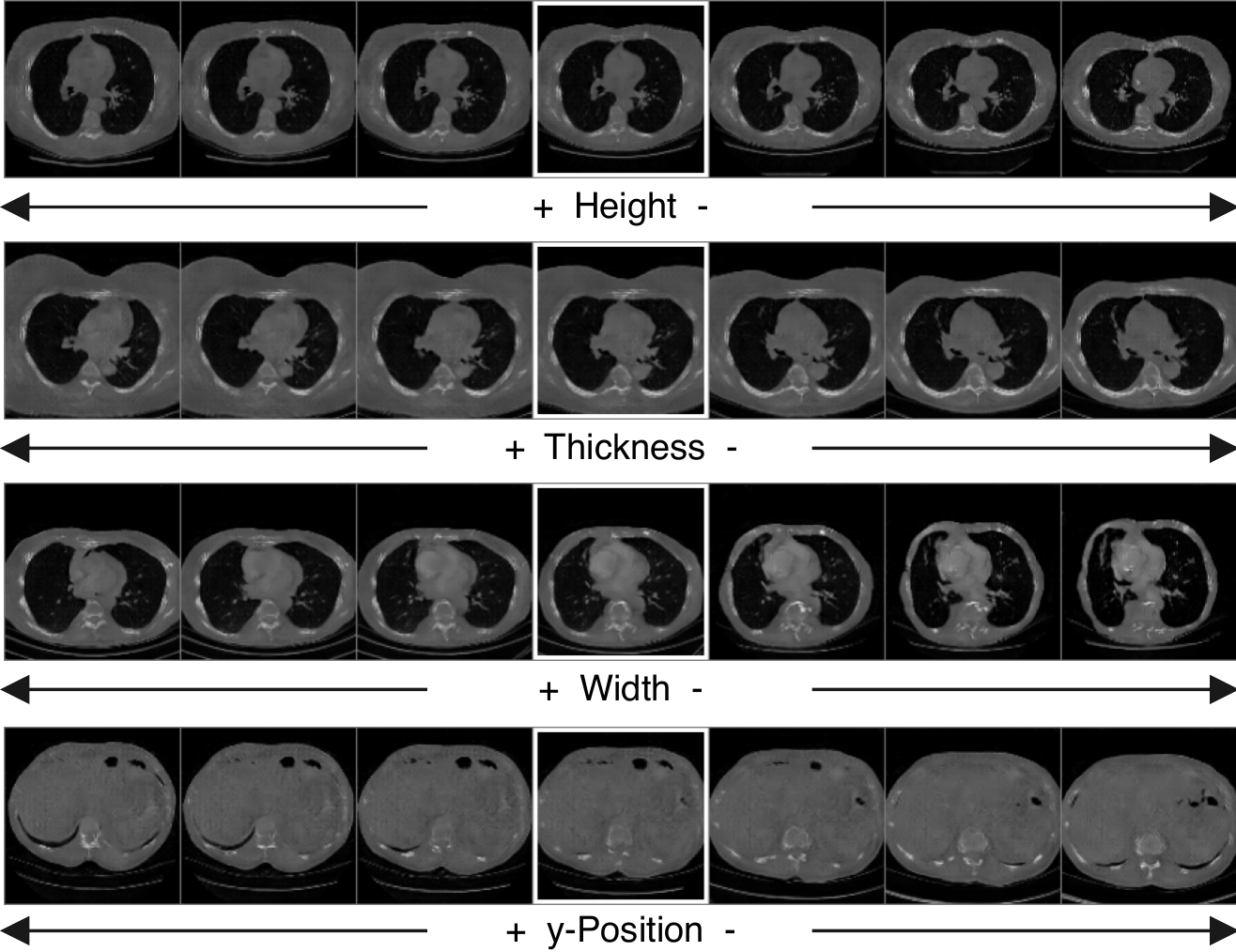}
    
  }
  \caption{Example of interpretable directions using $A^{32\times 32}$ with orthonormal columns, LeNet as reconstructor, the \ac{VAE} (a) and \ac{GAN} (b) as underlying generative models. The central images correspond to the original latent vector. The images to the left/right of that correspond to a negative/positive shift. We observed fewer disentangled directions than with other methods.}
\end{figure}

\begin{figure}[h!]
\centering
  {%
    \subfigure{%
      \label{fig:VaeResOrtho}
      \includegraphics[width=0.4\linewidth]{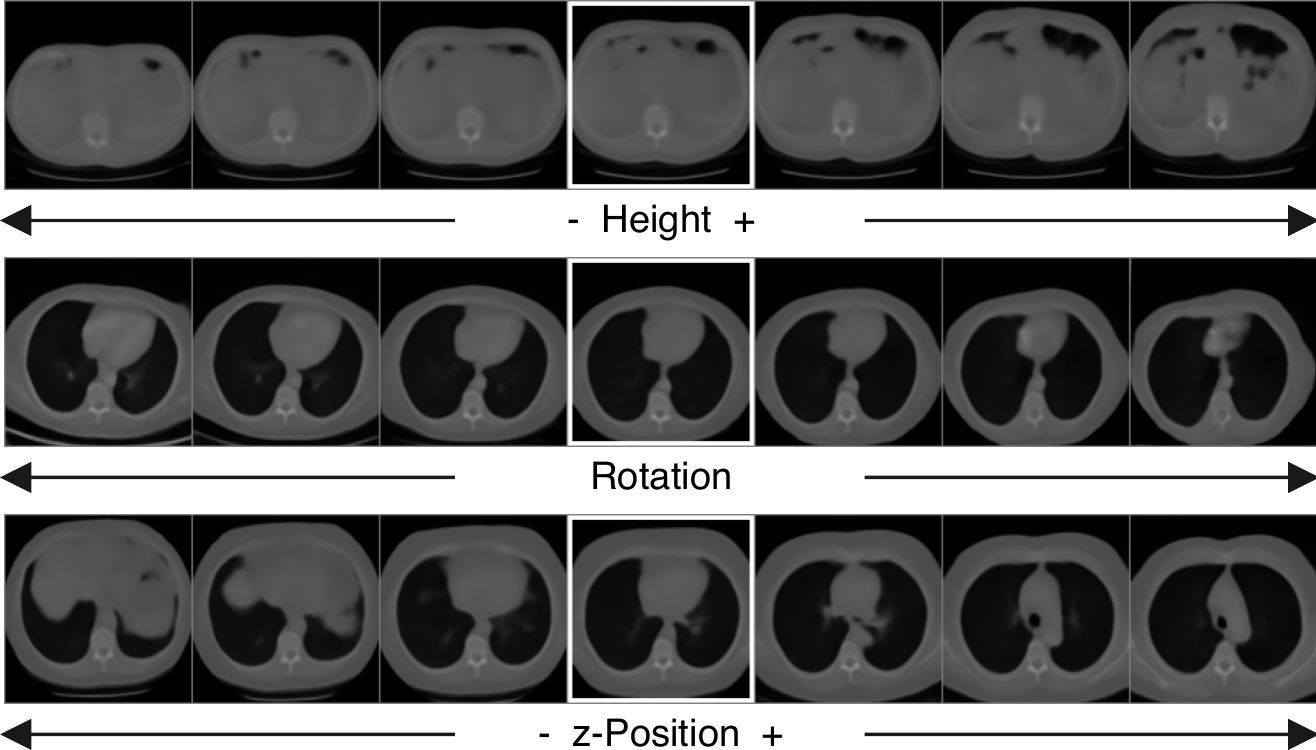}
    }\qquad 
    \subfigure{%
      \label{fig:GanResOrtho}
      \includegraphics[width=0.4\linewidth]{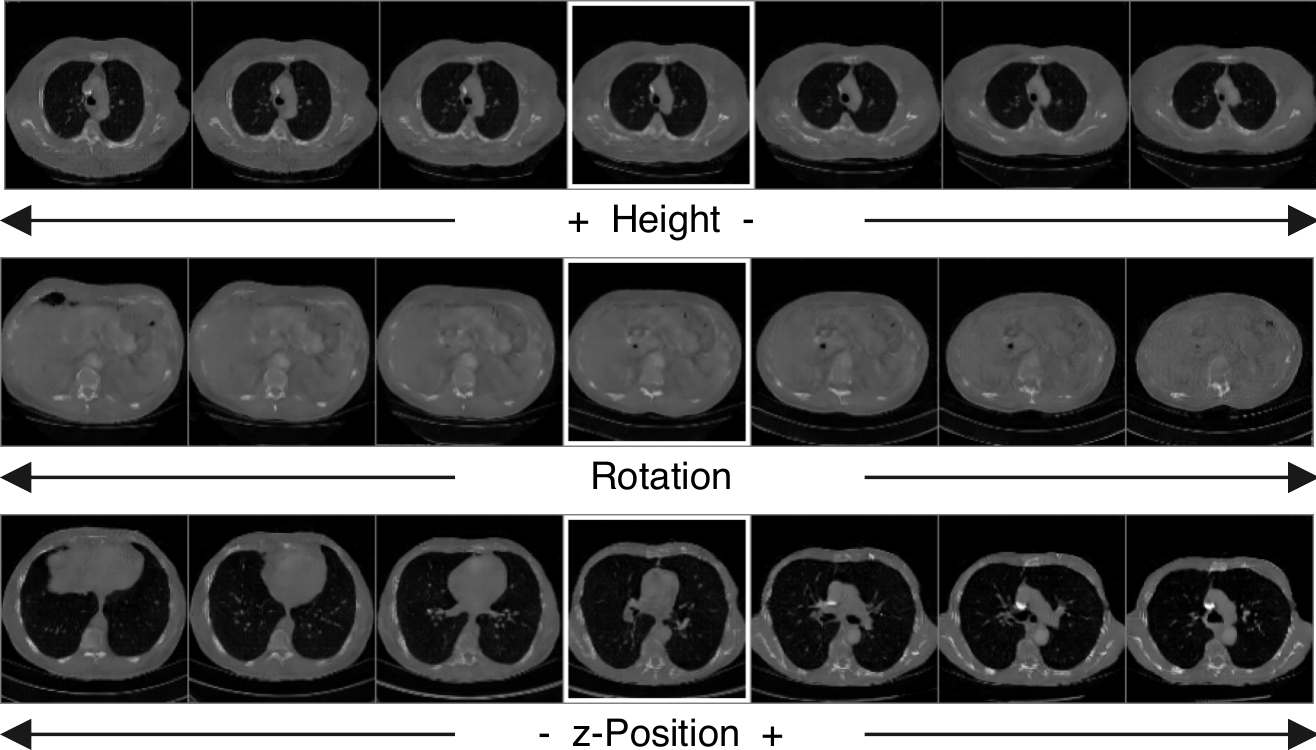}
    }
  }
  {\caption{Example of interpretable directions using $A^{32\times 32}$ with orthonormal columns, ResNet as reconstructor, the \ac{VAE} (a) and \ac{GAN} (b) as underlying generative models. The central images correspond to the original latent vector. The images to the left/right of that correspond to a negative/positive shift. Again, we observe far fewer disentangled directions compared to the other methods limiting the amount of directions we report.}}
\end{figure}
\begin{figure}[h]
\centering
  {%
    \subfigure{%
      \label{fig:VaeLeProj}
      \includegraphics[width=0.39\linewidth]{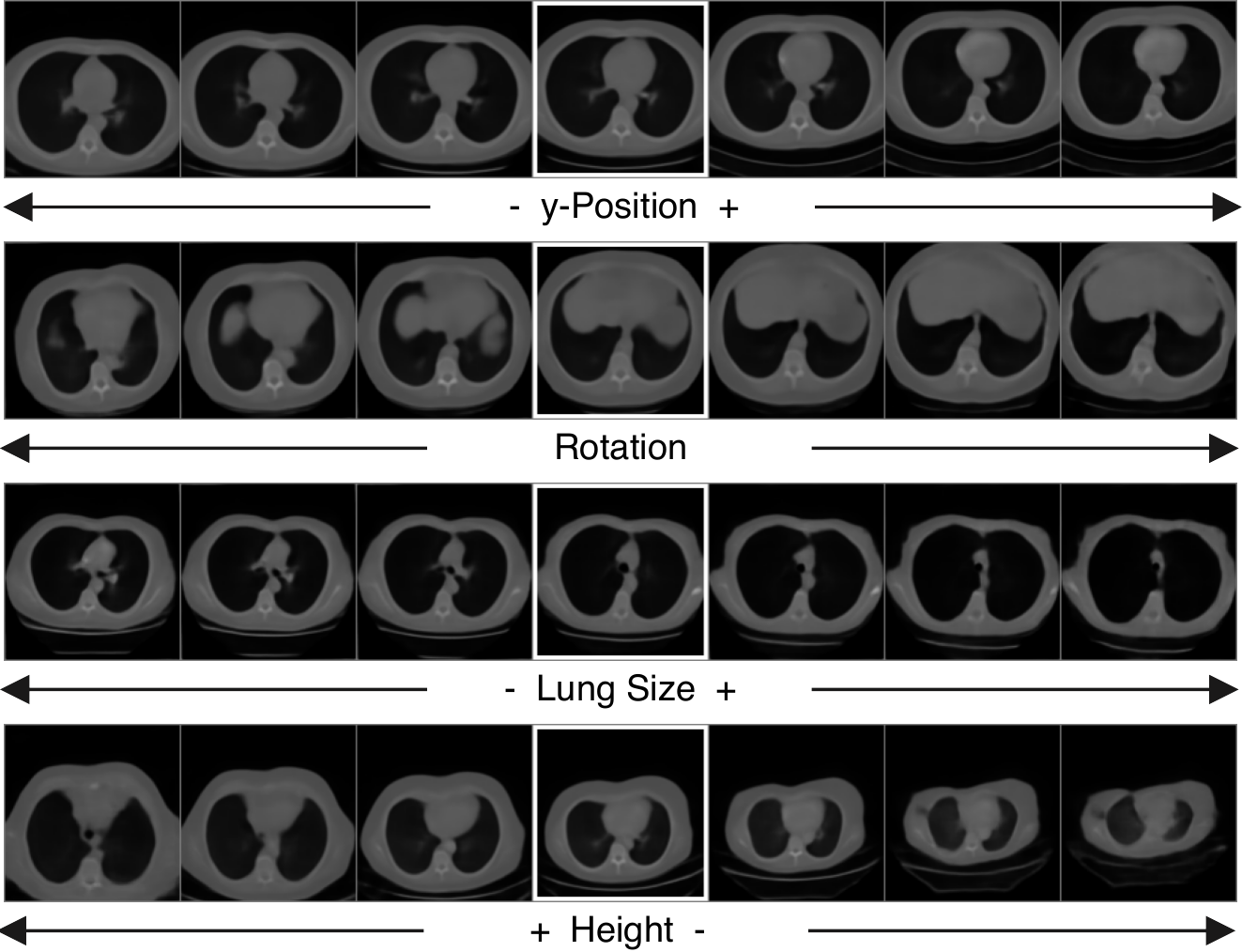}
    }\qquad 
    \subfigure{%
      \label{fig:GanLeProj}
      \includegraphics[width=0.39\linewidth]{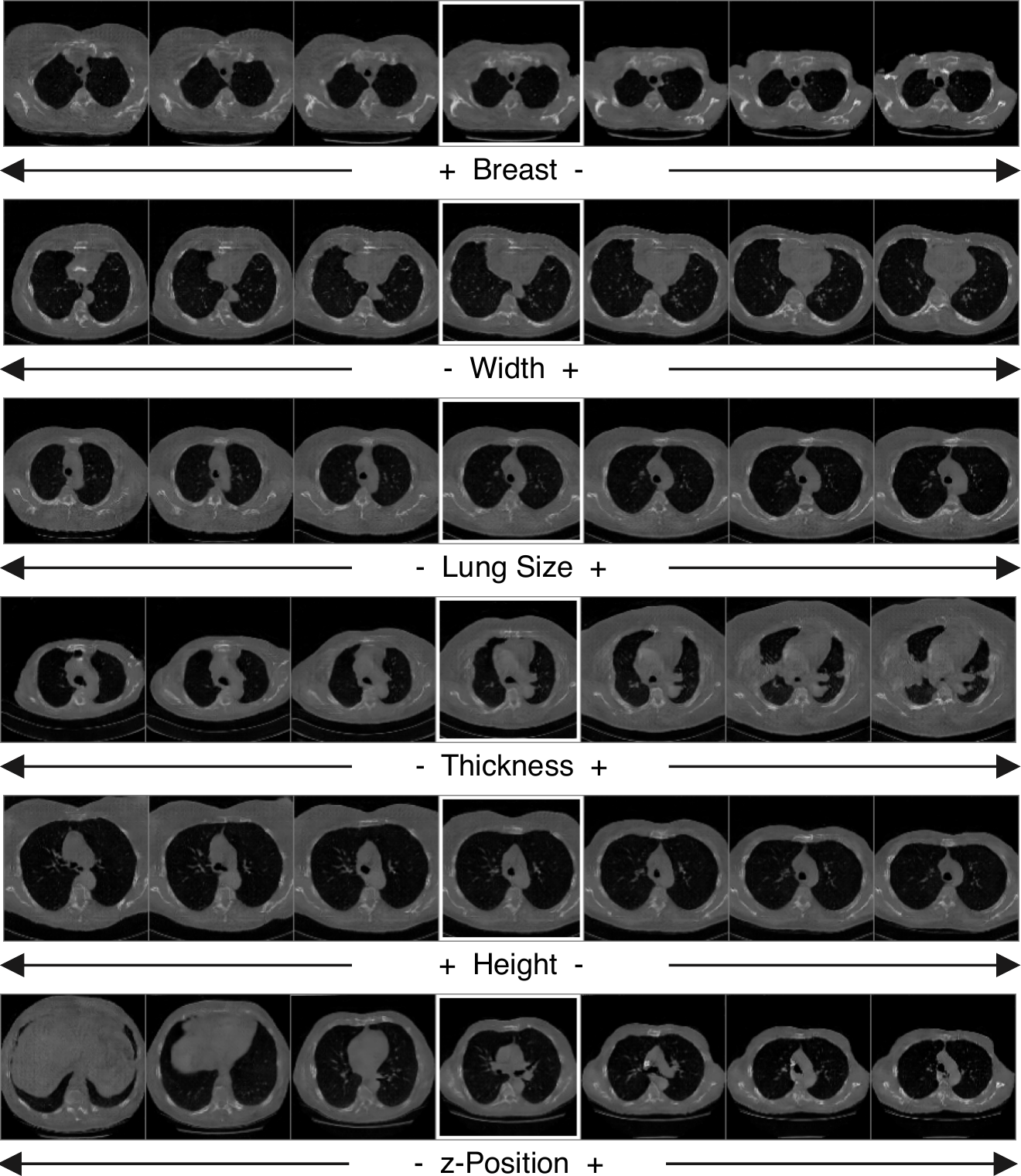}
    }
  }
  {\caption{Example of interpretable directions using $A^{32\times 32}$ with columns of unit length, LeNet as reconstructor, the \ac{VAE} (a) and \ac{GAN} (b) as underlying generative models. The central images correspond to the original latent vector. The images to the left/right of that correspond to a negative/positive shift.}}
\end{figure}
\begin{figure}[h]
\centering
  {%
    \subfigure{%
      \label{fig:VaeResProj}
      \includegraphics[width=0.4\linewidth]{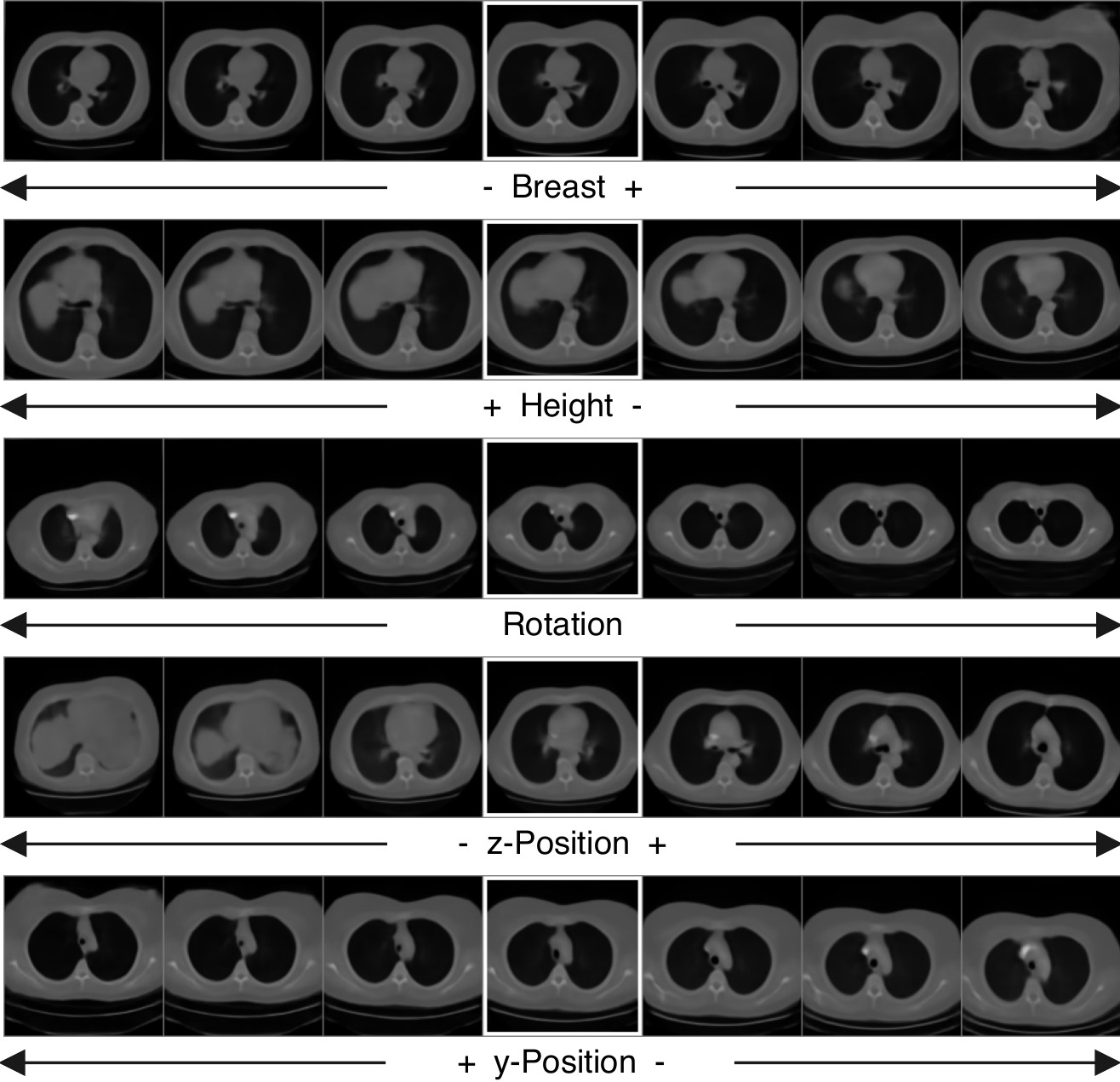}
    }\qquad 
    \subfigure{%
      \label{fig:GanResProj}
      \includegraphics[width=0.4\linewidth]{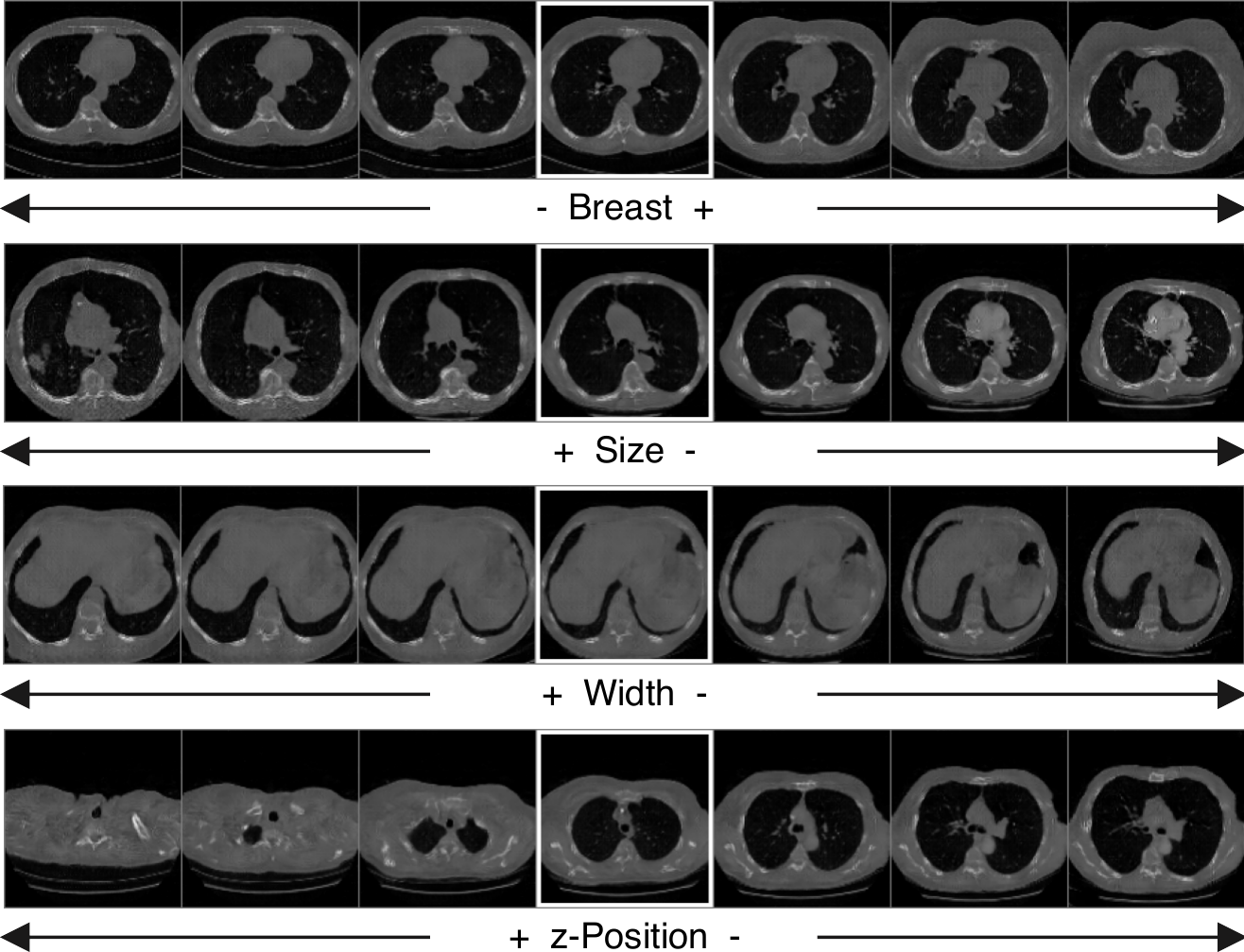}
    }
  }
  {\caption{Example of interpretable directions using $A^{32\times 32}$ with columns of unit length, ResNet as reconstructor, the \ac{VAE} (a) and \ac{GAN} (b) as underlying generative models. The central images correspond to the original latent vector. The images to the left/right of that correspond to a negative/positive shift.}}
\end{figure}
\begin{figure}[h]
\centering
  {%
    \subfigure{%
      \label{fig:VaeRes100}
      \includegraphics[width=0.4\linewidth]{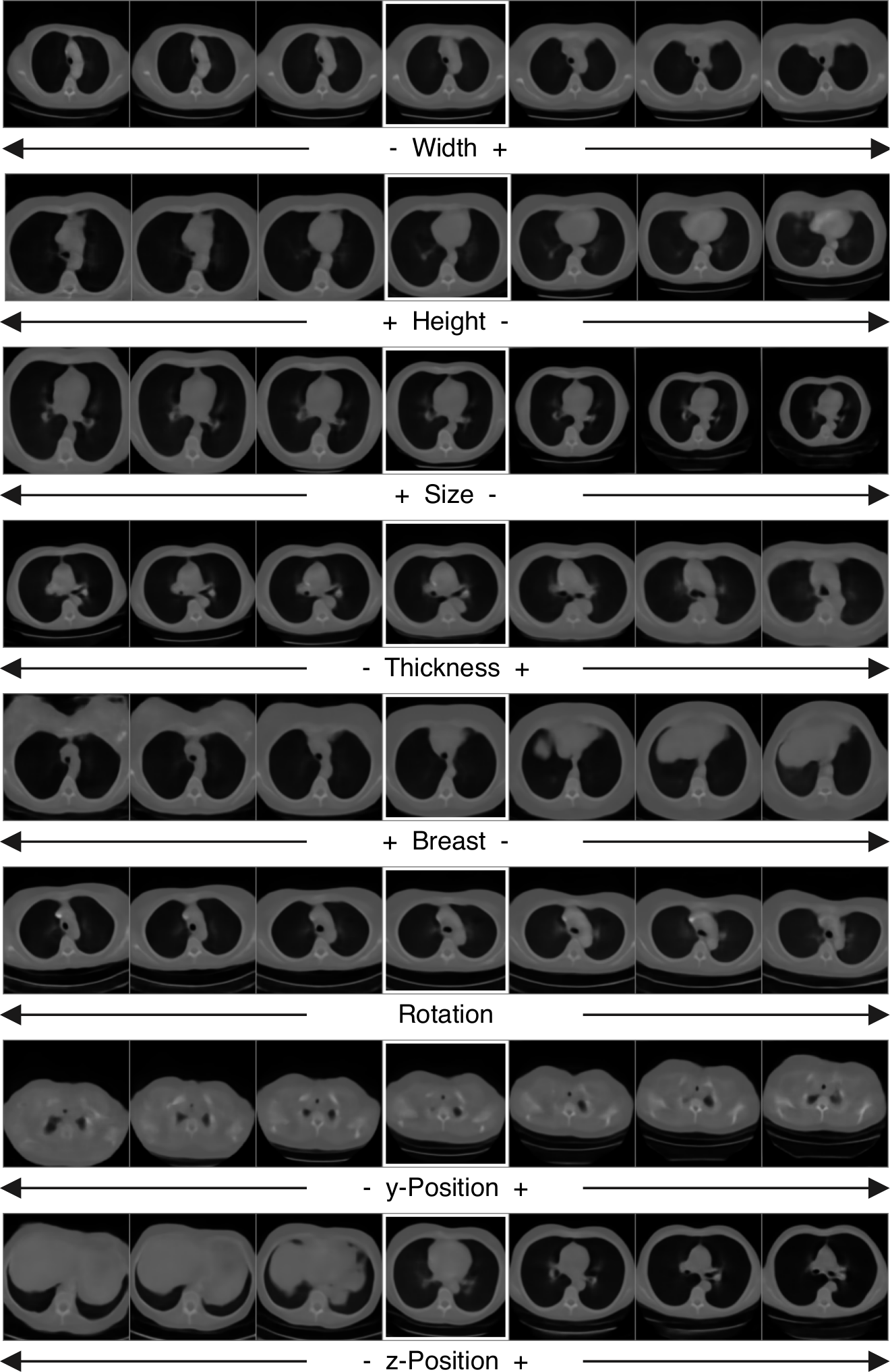}
    }\qquad 
    \subfigure{%
      \label{fig:GanRes100}
      \includegraphics[width=0.4\linewidth]{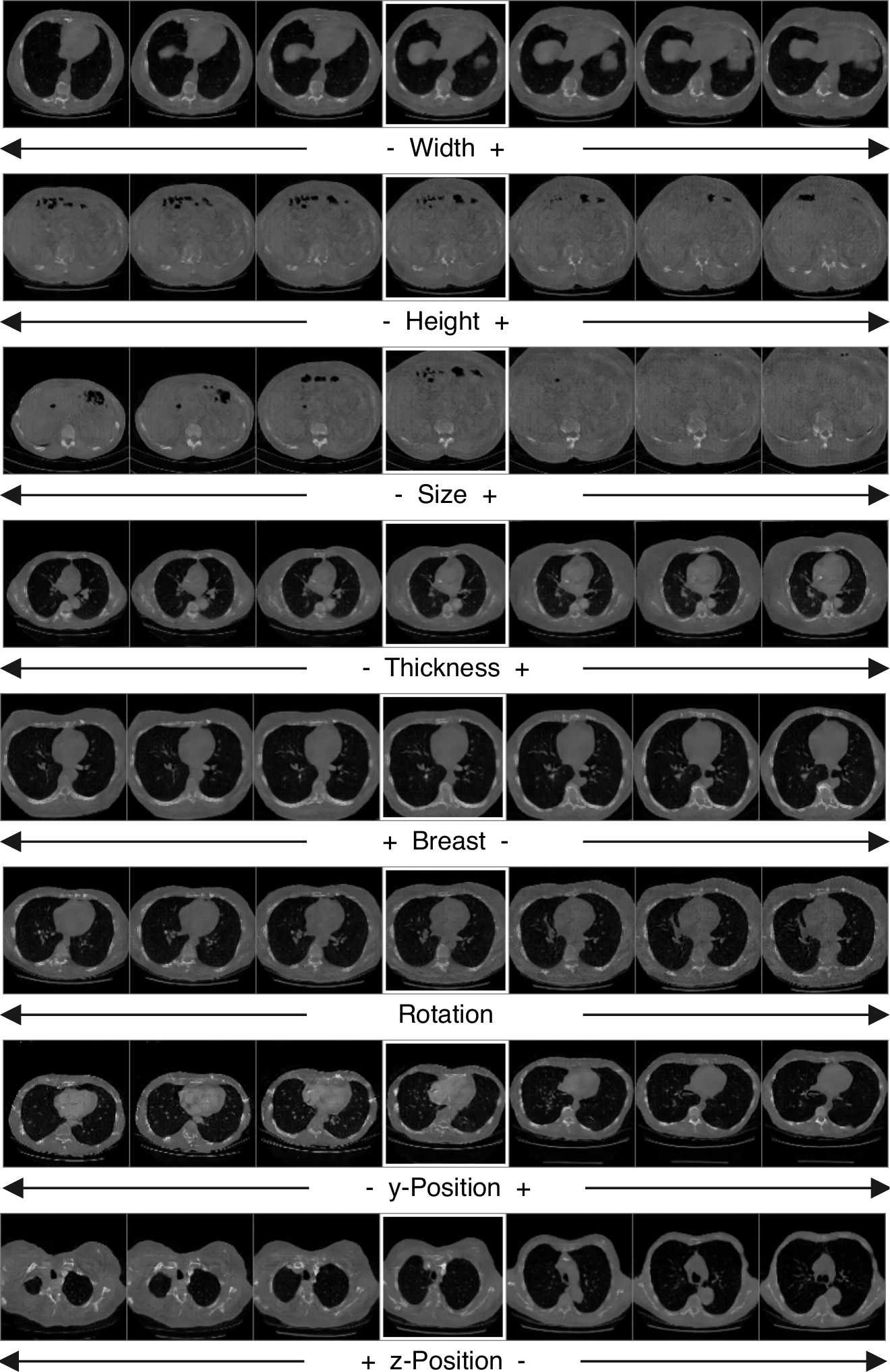}
    }
  }
  {\caption{Example of interpretable directions using $A^{32\times 100}$ with columns of unit length, ResNet as reconstructor, the \ac{VAE} (a) and \ac{GAN} (b) as underlying generative models. The central images correspond to the original latent vector. The images to the left/right of that correspond to a negative/positive shift.}}
\end{figure}

\FloatBarrier
\section{Supplementary Images}
\label{app:more_images}
\begin{figure}[h]
\centering
  {\includegraphics[width=0.79\linewidth]{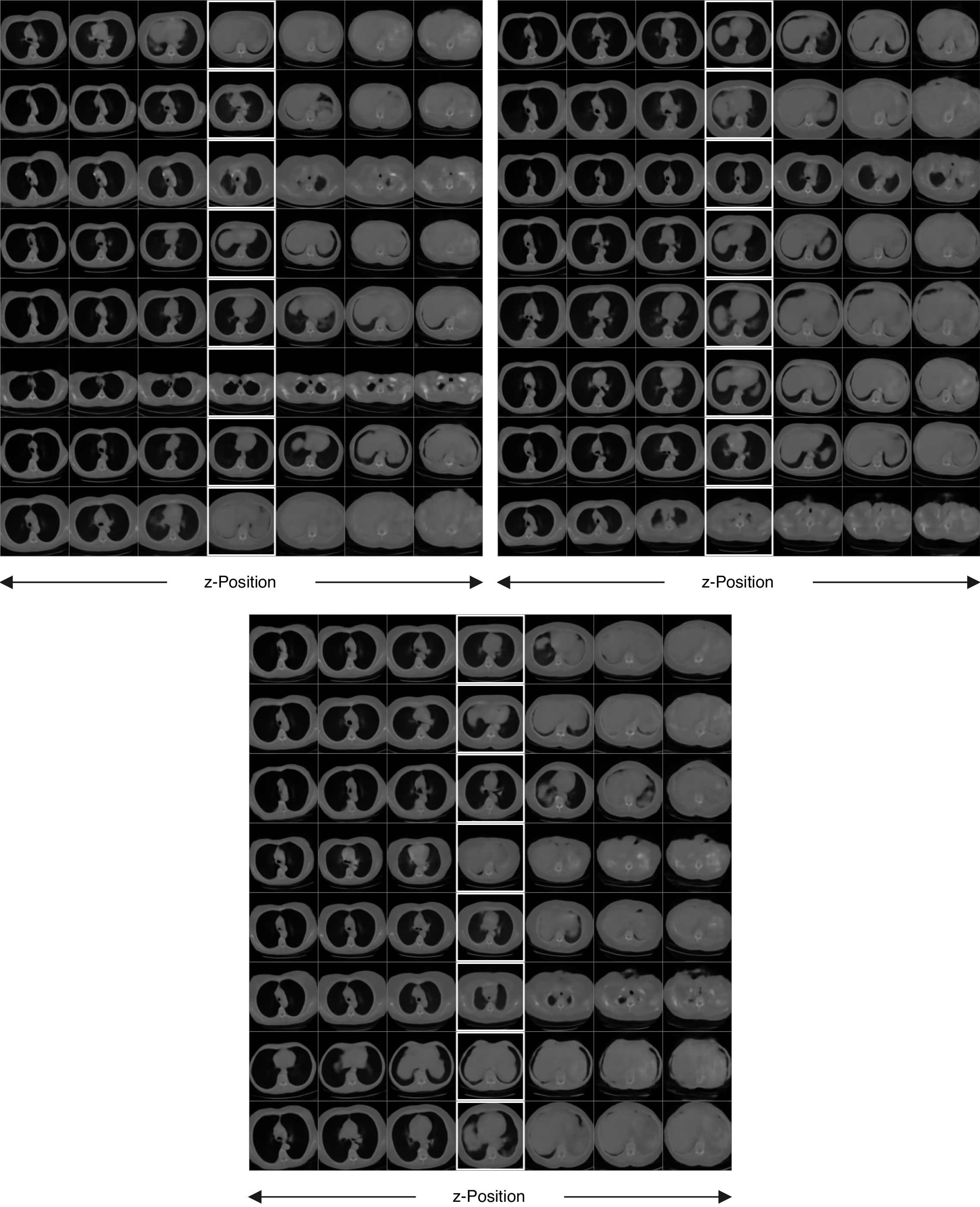}}
  {\caption{24 Randomly sampled latent vectors shifted along the directions corresponding to $z$-Position. The central images correspond to the original latent vector. The images to the left/right of that correspond to a negative/positive shift. Each latent vector shows biological variation and all shifts show realistic changes in anatomy corresponding to different z-positions in order of the amount of shift, such as different anatomical areas of the airways, heart, lungs, and liver.}}
\end{figure}

\begin{figure}[h]
\centering
  {\includegraphics[width=\linewidth]{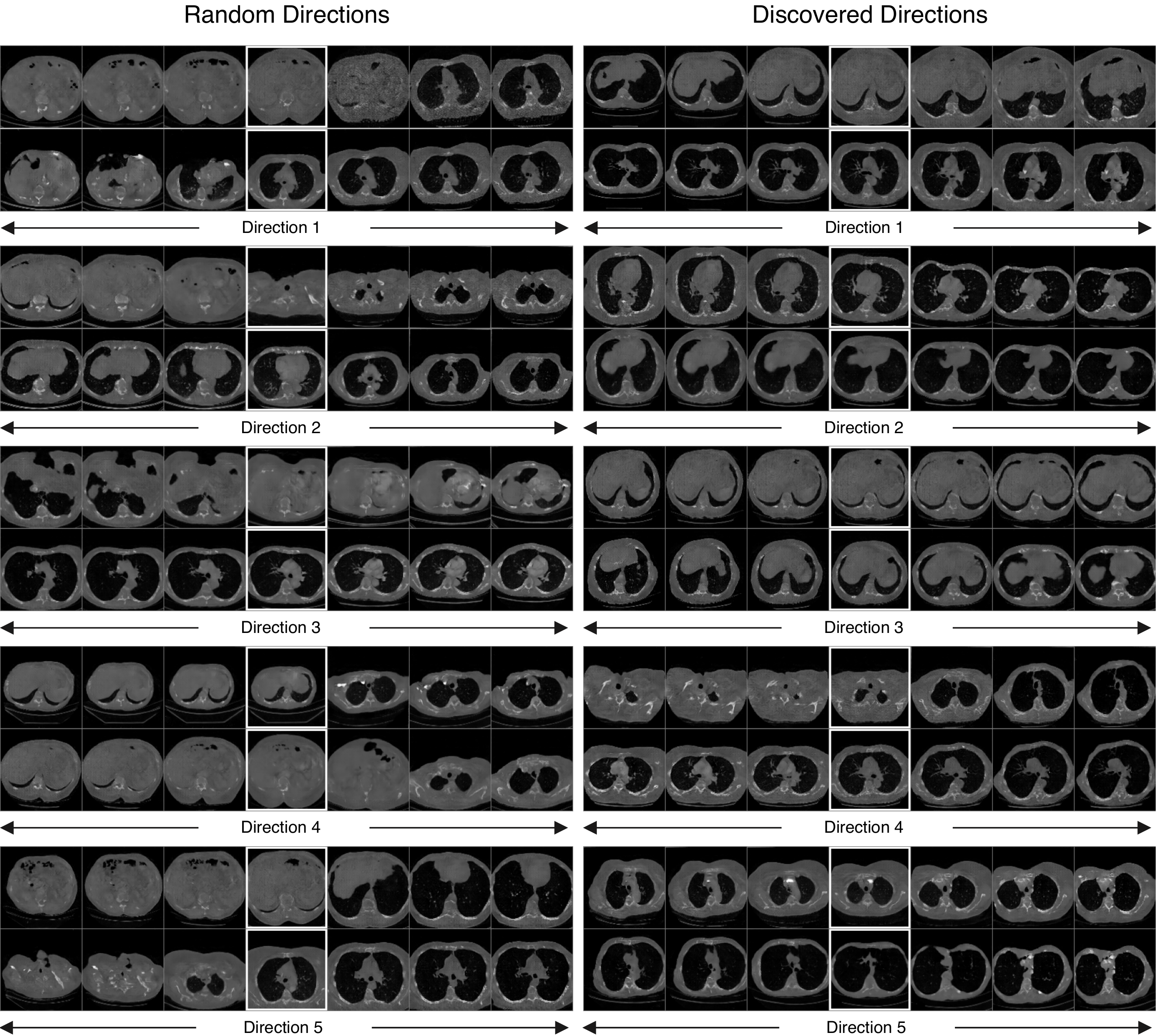}}
  {\caption{5 Random directions (left) next to the first 5 discovered directions (right) using $A^{32\times 100}$ with columns of unit length, LeNet as reconstructor and the \ac{GAN} as underlying generative model. The central images correspond to the original latent vector. The images to the left/right of that correspond to a negative/positive shift. The results show a marked difference in interpretability of the discovered directions in contrast to random directions.}}
\end{figure}
\begin{figure}[h]
\centering
  {\includegraphics[width=0.9\linewidth]{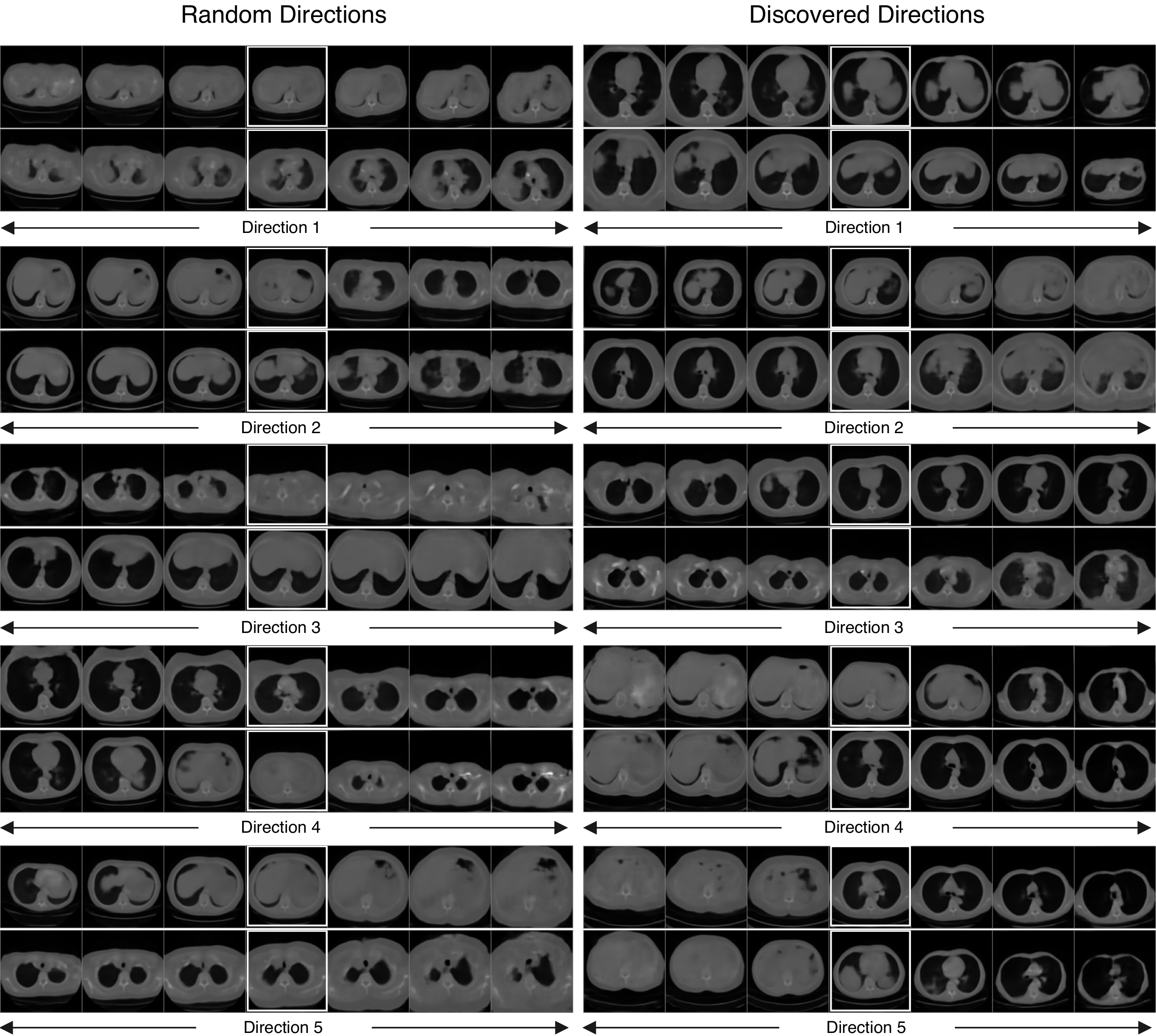}}
  {\caption{5 Random directions (left) next to the first 5 discovered directions (right) using $A^{32\times 100}$ with columns of unit length, LeNet as reconstructor and the \ac{VAE} as underlying generative model. The central images correspond to the original latent vector. The images to the left/right of that correspond to a negative/positive shift. We can see that the random directions are continues and some of them somewhat interpretable. In particular, the random direction results are better than what we observe with the \ac{GAN}. This is most likely due to the regularization of the latent space as well as the \ac{VAE} learning a structured latent space which is in contrast to the \ac{GAN}.}}
\end{figure}

\FloatBarrier
\section{Memorization}
\label{app:mem}
\begin{figure}[h]
\centering
  {\includegraphics[width=0.6\linewidth]{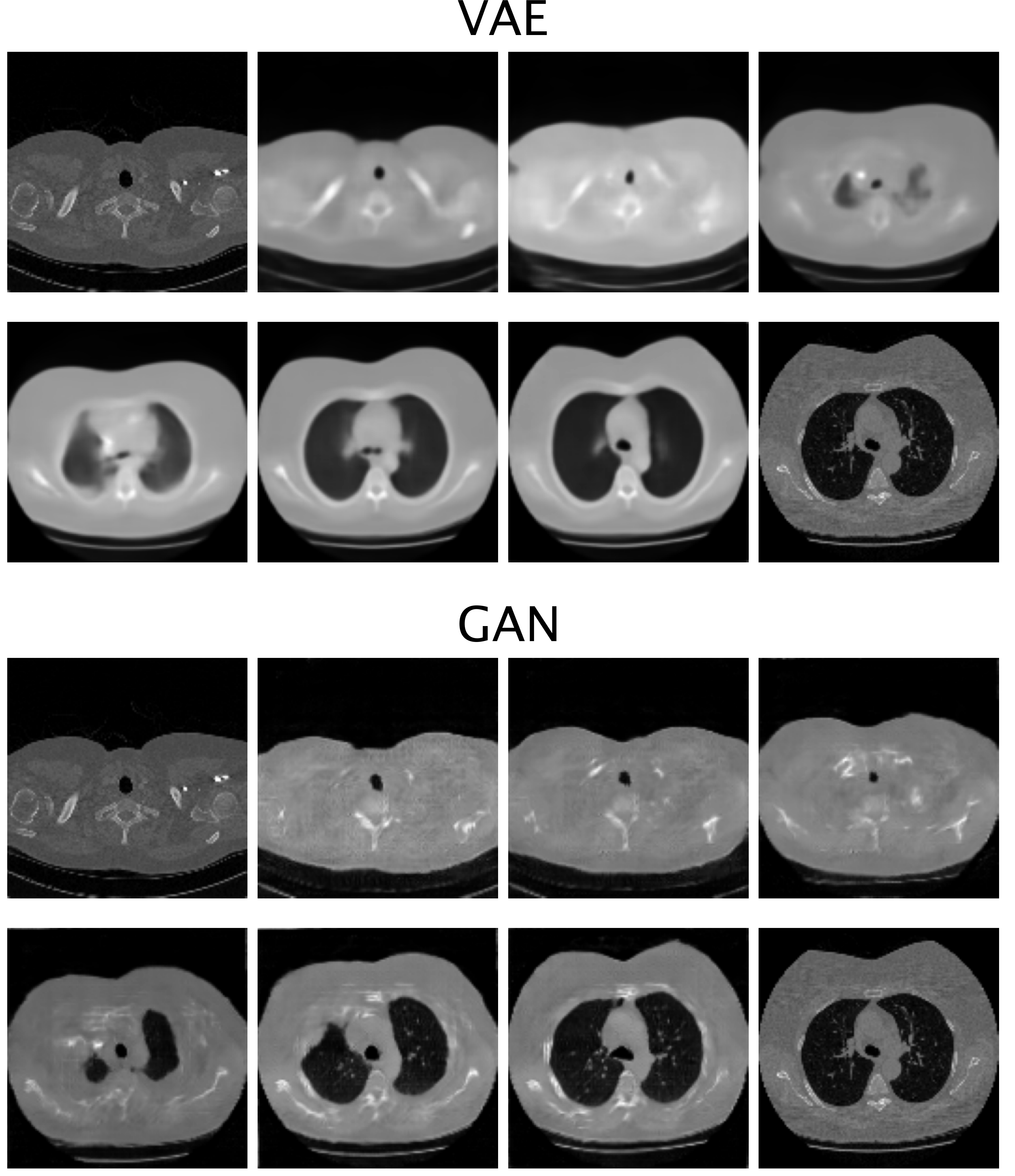}}
  {\caption{When considering generative latent models, memorization is a concern. To address this, we embed two random images from hold out data (original images are in the top left and bottom right) using reverse latent vector search \cite{Blanco21} for the GAN and the encoder for the VAE and move between the two resulting latent vectors to investigate the smoothness of the latent space. As can be seen, the resulting images appear anatomically correct, and interpolation is smooth, indicating that the generative models are not memorizing the data.}}
\end{figure}

\begin{acronym}
 \acro{RCA}{Reconstructor Classification Accuracy}
 \acro{KLD}{Kullback-Leibler Divergence}
 \acro{ELBO}{Evidence Lower Bound}
 \acro{VAE}{Variational Autoencoder}
 \acrodefplural{VAE}{Variational Autoencoders}
 \acro{GAN}{Generative Adversarial Network}
 \acrodefplural{GAN}{Generative Adversarial Networks}
 \acro{DCGAN}{Deep Convolutional Generative Adversarial Network}
 \acrodefplural{DCGAN}{Deep Convolutional Generative Adversarial Networks}
 \acro{LIDC-IDRI}{The Lung Image Database Consortium image collection}
 \acro{TCIA}{The Cancer Imaging Archive}
 \acro{CT}{Computed Tomography}
 \acro{HU}{Hounsfield Scale}
 \acro{DICOM}{Digital Imaging and Communications in Medicine}
 \acro{FID}{Fréchet Inception Distance}
\end{acronym}

\end{document}